# An Interactive Polymer Building Toolkit for Molecular Dynamics Simulations: PolyMAPS

*Xiaoli Yan*[*,1], *Santanu Chaudhuri*[1,2]

[1]Department of Civil, Materials, and Environmental Engineering, University of Illinois at Chicago, 815 West Van Buren St, Suite 525, Chicago, IL 60607, United States

[2]Also at: Applied Materials, and Data Science and Learning Division, Argonne National Laboratory, Lemont, Illinois 60439, USA

*Author to whom correspondence should be addressed: xyan11@uic.edu

ABSTRACT

PolyMAPS is an open-source library that helps researchers to initialize LAMMPS molecular dynamics simulations. It introduces an integrated workflow by combining preparation, launching, visualization, and analysis into a single Jupyter notebook. PolyMAPS enables users to build small or polymeric molecules in a user-friendly interactive 3D plotting system that supports reading and writing systems in LAMMPS data file format. Hence, PolyMAPS demonstrates the potential of reducing the learning difficulties of new users of the LAMMPS software.

## Introduction

Many open-source preprocessing tools are developed for molecular dynamics (MD) simulations for simulation software such as LAMMPS[1], GROMACS[2,3], CP2K[4], etc. Analysis and building tools, such as Moltemplate[5], Packmol[6], VMD[7], EMC[8], OVITO[9], Avogadro[10], PyLAT[11], etc., are mostly designed by experienced scientists who are very familiar with simulation workflow. These are all excellent software packages that specialize in their own domains, and they are made freely available to the scientific community by diligent scientists and engineers. However, a reasonably complex simulation system is usually built, simulated, and analyzed by multiple different pieces of open-source software that require various styles of user interface and experience. Commercial software packages with pre-designed workflows, such as Scienomics MAPS[12], Materials Studio[13], MedeA[14] etc., can sometimes simplify the process. These issues could lead to a steep learning curve and entry barriers for the students and beginner scientists who are relatively new to this knowledge domain. For the entry-level audiences in the field of MD simulations, an intuitive and graphical interface would be immensely helpful. A unified programming interface is helpful for the users to develop customized functionalities in the long run.

Input and output file formats of different simulation software vary considerably. In many situations, the same piece of information describing the same molecular system appears in various file formats during the prototyping stage of a simulation, e.g.: LAMMPS uses its own specialized and customizable file formats for static systems and trajectories; XYZ file can contain static systems and trajectories that contains atoms but not bonds; PDB file can record static systems and trajectories with both atoms and bonds, but it has certain limitations since it was developed in the 1970s. In addition, many old file formats have redundant information due to compatibility issues. Software packages, like OpenBabel[15] and RDKit[16], become convenient when files are being converted from one format into another. On the other hand, the Simplified Molecular-Input Line-Entry System (SMILES)[17–19] has become popular in recording small molecules in the cheminformatics community since it is portable and readable to both humans and machines[20–22]. SMILES notation and its variants have also been employed in the field of polymeric science given its capability and flexibility in describing molecular structures with linear and branched topologies. BigSMILES[23] has been avant-garde in exploring the possibility of abstracting compositional information with specialized SMILES notations and syntax.

Many open-source visualization packages excel in their own domain expertise, such as VMD[7] and OVITO[9], but a similar learning barrier exists here. VMD has both Tcl, Python command interfaces, and a well-organized GUI, but not all functionalities are available through GUI. OVITO has a Python command interface and a comprehensive GUI menu with functionalities well-documented on its website, and it even supports querying information via mouse hovering over atoms. The visualization windows of both programs are clickable to manipulate the view angle and zoom levels for molecular systems. VMD's GUI and command line are both free, but OVITO's GUI requires a license when the Python scripting module is involved. Another noticeable effort that has similarities to this work is Polymer Structure Predictor (PSP)[24]. It also has a Jupyter-based interface and in-place visualization, and its visualization is based Py3Dmol[25], a Python-interfaced 3Dmol.js. Although being an interactive and portable 3D visualization library, Py3Dmol lacks the capability of a tooltip on mouse hover for in-place information querying. A fancier approach is the Game-Engine-Assisted Research platform for Scientific computing (GEARS)[26], which leverages the existing graphics and physics technology from popular game engines, like Unity[27] and Unreal[28], to analyze and visualize LAMMPS simulations. The user interface to input commands to the system via virtual reality headsets and controllers offers the users an immersive experience that is albeit a great advancement, it is beyond the current status of the average working atmosphere in a classroom or an office.

For an average MD simulation, a typical workflow can be defined with the following steps:

1. building single molecule (or monomer) geometry and topology,
2. force field parameterization and assignment,
3. generating a polymeric or bulk or interfaced system for studying a group of specific properties,
4. launching simulation software suite with computational resource allocation,
5. generating plots and tables from the captured outputs of the simulation.

At each step, the user should have as much access to visual inspection and in-place information queries as possible to minimize the effort in prototyping. Pysimm[29] is a Python-based LAMMPS simulation control API for polymer building and simulations. It enables user to unify the whole workflow in Python from building to simulating polymeric and small molecular systems. Pysimm supports many force fields: CHARMM[30], DREIDING[31], AMBER[32–34], and PCFF[35]. RadonPy[36] is a recently published library that automates the property calculations with the AMBER force field. Both Pysimm and RadonPy are great contributions in the field of polymer simulations, especially in the sense of automation of property calculations with predefined simulation protocols. This enable the high-throughput calculation that curates the RadonPy's property database[37]. stk[38] is a Python library that focuses on generating molecular and supramolecular structures. It even has a custom GUI and database for structure data visualization and storage. PySoftK[39] is also a recently published Python library that focuses on enabling user to generate a variety of polymer structures.

This work introduces a generalized and lightweight workflow, Polymeric Molecule Assembly Programming System (PolyMAPS), for prototyping and analyzing polymeric systems. The essence of designing a comprehensive workflow for MD simulations is to integrate as many functionalities as possible into one software environment as possible. Thus, by significantly reducing the learning barrier of programming and different types of pre-processing and post-processing software, the PolyMAPS workflow is designed to initiate a LAMMPS simulation for testing and demonstration purposes.

There are certain limitations in each part of the software's functionalities due to the computational and graphics capabilities of a Jupyter notebook[40]. The input and output system could be designed to leverage the portability of SMILES notation and the robustness of native file formats supported by LAMMPS. Simulation systems with a very large number (>100,000) of atoms are out of the scope of this work.

## Methodology

The center of the PolyMAPS workflow is a Jupyter notebook with Python kernel. As a general-purpose programming language, Python has been widely welcomed in the data science community. Instructions written in Python using NumPy[41], SciPy[42], and Pandas[43] packages are implemented to manipulate the coordinates and other parameters of atoms and molecules. The workflow can be visualized in Figure 1.

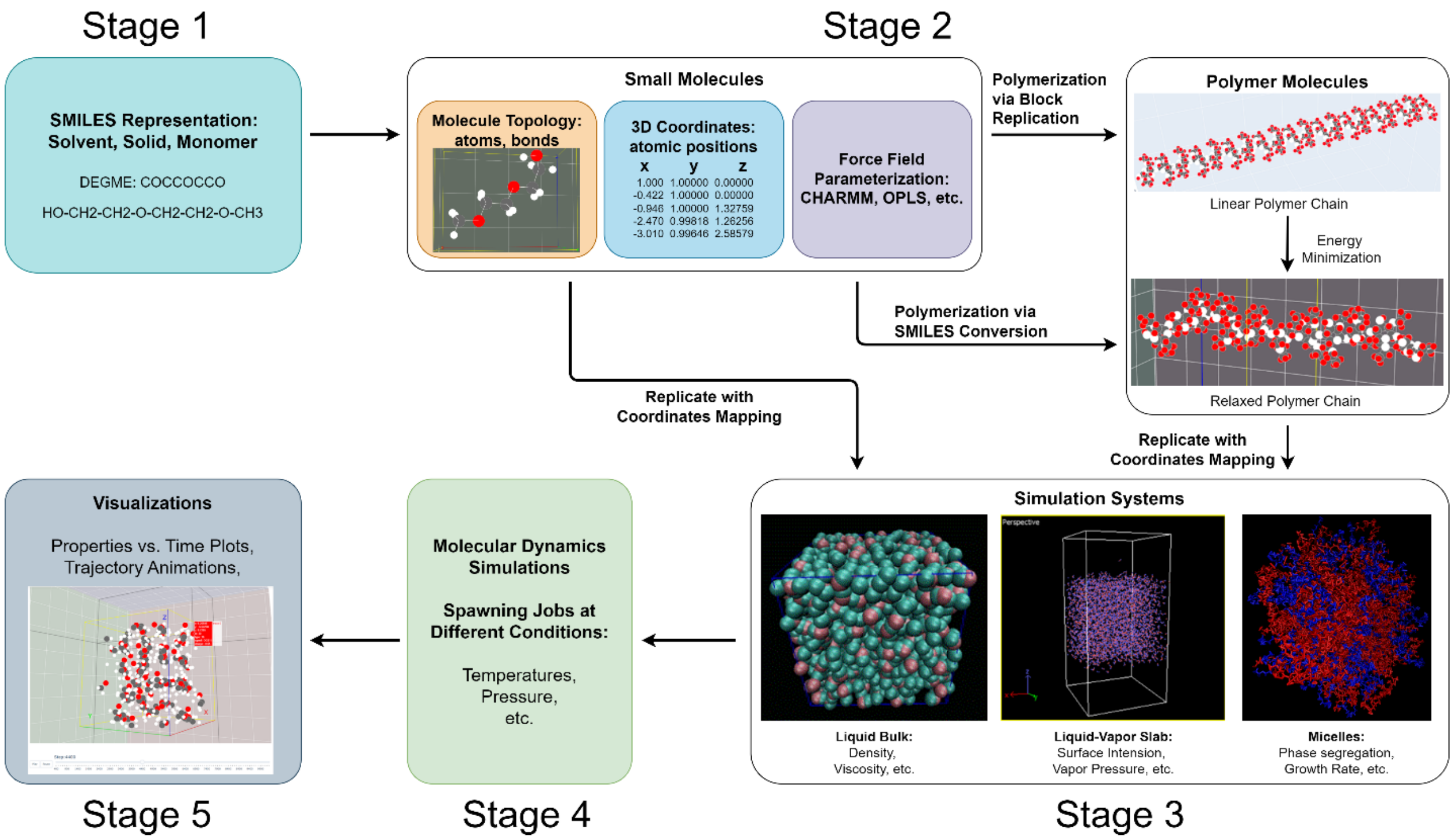

**Figure 1.** Molecular simulation building workflow with five stages of operations. Arrows denote the data flow directions. 3D visualizations are generated by OVITO, VMD, and PolyMAPS.

An interactive visualization tool is developed based on Plot.ly[44] to demonstrate the atoms and bonds with their corresponding indices and atom information. User can query the related information about a specific atom by hovering the mouse cursor over the atom in the 3D structure visualization. Information such as: xyz coordinate, partial charge, atom id, atom type, force field type, etc. can all be queried with minimal effort. It is our opinion that this functionality separates PolyMAPS from all existing 3D visualization tools and provides users with unprecedented convenience to the users. This functionality will be demonstrated in the description of the usage cases.

In stage 1, the program starts with either a monomer segment or a complete molecule in the LAMMPS data format. If a complete molecule is already defined, stage 1 can be skipped. A linear polymer chain will require a copy of the left terminal segment, a copy of the right termination segment, and copies of the monomers forming the linear polymer chain. Typically, a termination segment has one linkage anchor, and a monomer segment has two linkage anchors. Branched polymeric systems can also be formed by defining monomers with more than 2 linkage anchors. With a user-defined degree of polymerization, the program can generate a linear polymer chain with 3D coordinates and the correct bond linkage between monomer segments. The built polymer chain is packaged as a molecule in LAMMPS data format for further processing. The atomic partial charge and atom types are also determined in this stage.

In stage 2, the force field parameters are assigned to the molecule that was built in stage 1. For example, the CHARMM general force field (CGenFF)[30] files are parsed, and the related parameters are loaded into Pandas data frames by screening all the atom types. By matching all the possible permutations of the many-body interactions from the loaded parameter data frames: angle, dihedral, and improper, all interactions are detected and assigned to the corresponding selection of atoms automatically. OPLS-AA force field[45] is also supported by directly using an externally generated LAMMPS data file from the LigParGen website[46–48]. Then, the workflow can skip stages 1 and 2, and go straight to stage 3.

In stage 3, a larger-scale system that is statistically meaningful can be built using the defined molecules. Packmol is commonly used in the community in order to generate randomly initialized systems for MD simulations. However, the supported file formats are limited to XYZ, PDB, and Tinker. For XYZ, no bonding information is present before or after the coordinates' manipulation; for PDB and Tinker, bonding information can be defined and preserved through the process but converting the Packmol output file to LAMMPS data format with angles, dihedrals, impropers and partial charges could be challenging. PolyMAPS offers the user the opportunity to choose how to manipulate atomic coordinates by separating this process from the simulation initialization workflow. Users can choose whether to use Packmol or to generate new atomic coordinates within PolyMAPS or with LAMMPS built-in functions. A single molecule system with assigned force field parameters and box boundaries will act as a building block to prepare a more complicated system for computational studies. By duplicating the existing building blocks and stacking them into a larger system using user-defined box sizes, a bulk polymer system can be built. This is achieved by invoking the "replicate" command from LAMMPS. Mixing different types of molecules is also supported for complex systems such as aqueous solutions, multi-phase segregation systems, multi-phase dispersion systems, etc. An example of water-isopropanol mixture is included in the Supporting Information case 6.

In stage 4, users can define the LAMMPS simulation condition with either a text file-based command interface or the PyLAMMPS[49] interface within the Python environment. For small and simple systems, users can launch jobs locally, but for larger and complex systems that require computational cluster access, a job submission script generator is provided. Templates and sample scripts for the Torque[50] system and the Slurm[51] system are provided.

In stage 5, simulation result visualization can be achieved by using the interactive plotting capabilities of Plotly. A LAMMPS log parsing script is provided for the default log style. Selected properties of the system and computational performance metrics can be visualized in the same Jupyter notebook. In case a .xyz trajectory file is generated by LAMMPS, the workflow can help convert it to a regular .xyz trajectory with element symbols instead of atom type numbers. Using Py3Dmol[25], the trajectory animations can be visualized in the Jupyter notebook. An alternative trajectory visualization tool with interactive features is implemented using Plot.ly. Both the regular .xyz trajectory file format and customized LAMMPS trajectory dump format are supported.

Currently, PolyMAPS supports the following functionalities: step-by-step preparation of reactive simulation with non-reactive force field, homopolymer building, block copolymer building, micelle building and solvating, surface tension calculation, and viscosity calculation. 3 cases will be introduced here, and more descriptions about the functionalities are included in the Supporting Information.

## Results and discussion

*Case-1: Launching MD simulations inside a Jupyter notebook to generate polymers*

A polymeric structure with custom terminations, monomers, and polymerization schemes can be defined and generated using the PolyMAPS functionalities. The building process of a system with polyethylene oxide (PEO) molecules is demonstrated.

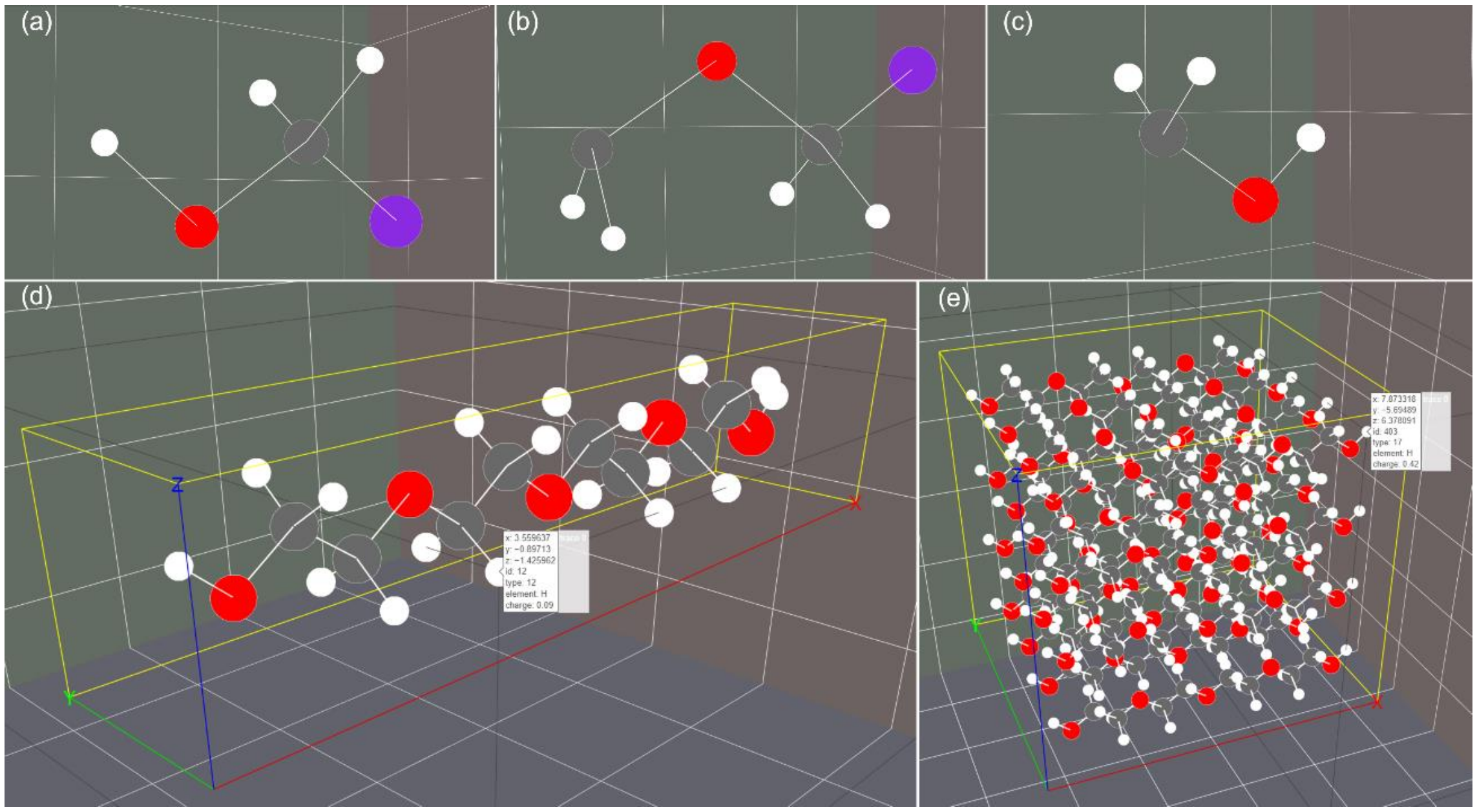

**Figure 2.** Building blocks of a polyethylene oxide molecule: (a) head termination block: OH-CH2-X with anchor position X (labeled with purple color); (b) monomer block: CH2-O-CH2-X with anchor position X (labeled with purple color); (c) tail termination block: CH2-OH; (d) assembled

PEO molecule; (e) a ready-to-simulate system with 16 PEO molecules with an orthogonal simulation box. Mouse-hovering tip enables immediate information query for any atoms in the 3D system. Atom color code: hydrogen as white, carbon as grey, oxygen as red.

As shown in Figure 2(d), a polyethylene oxide (PEO) with a degree of polymerization 4 can be decomposed into 2 termination blocks: OH-CH2-X, CH2-OH, and a repetitive monomer block: CH2-O-CH2-X, shown in Figure 2(a)(b)(c) respectively. Letter X is used to denote the anchoring position of the next building block's first atomic position. For example, when a monomer block is attached to a head block, the monomer block is first translated so that the left-most carbon atom overlaps with the anchoring position of the head block, site X from the head block is removed, and bonding information is updated.

Once a polymer molecule's atomic positions and bonding connectivities are all defined, the force field assignment process is ready to take place. PolyMAPS is designed to be force field agnostic so that the same molecule can be simulated with different flavors of force field parameterization schemes under a little amount of effort. Here, the assignment of CGenFF is demonstrated. Since the PEO molecule is documented in the force field files from CGenFF: top_all36_cgenff.rtf, the related atom types, bond types, angle types, dihedral types, and improper types are parsed and loaded into the Jupyter notebook as Pandas data frames, as shown below in Table S1~S4.

After the force field parameters are assigned to the molecule, the packaged molecule is stored in LAMMPS data format with atomic coordinates, topology definitions, and box boundaries. This LAMMPS data file is already syntactically correct and ready for simulation, but a system with many of the same molecules is needed. For example, one PEO polymer chain can be studied by running simulation on one polymer chain of varying length, but many polymer chain and their entanglement need to be prepared initially to simulation bulk PEO properties. A more complex system that is statistically meaningful can be generated by treating the built molecule with defined boundary box as a new level of building blocks. During this process, replications and linear transformations can be applied to the atomic coordinates while no force field parameter or bonding topology of the molecule is isolated and preserved. For instance, a 4-by-4-by-1 group of PEO molecules can be realized by utilizing the built-in command "replicate" from LAMMPS. The result of replication is visualized in Figure 2(e).

A relaxation simulation can be run on the polymer box with an energy minimization followed by an NPT ensemble simulation. The simulation temperature is set at 300 K, pressure at 1 atm, and time step at 1 fs/step. The minimization process ends in 500 steps, followed by 200,000 steps of an NPT ensemble. This run is conducted in a Google Colaboratory notebook with free CPU time and Conda[52] distributed LAMMPS. Later, a longer simulation of 11 million steps with 2x2x2 replication of the system with the same NPT settings was conducted to equilibrate the physical system. The temperature, pressure, density, and potential energy measurements are visualized in Figure 3.

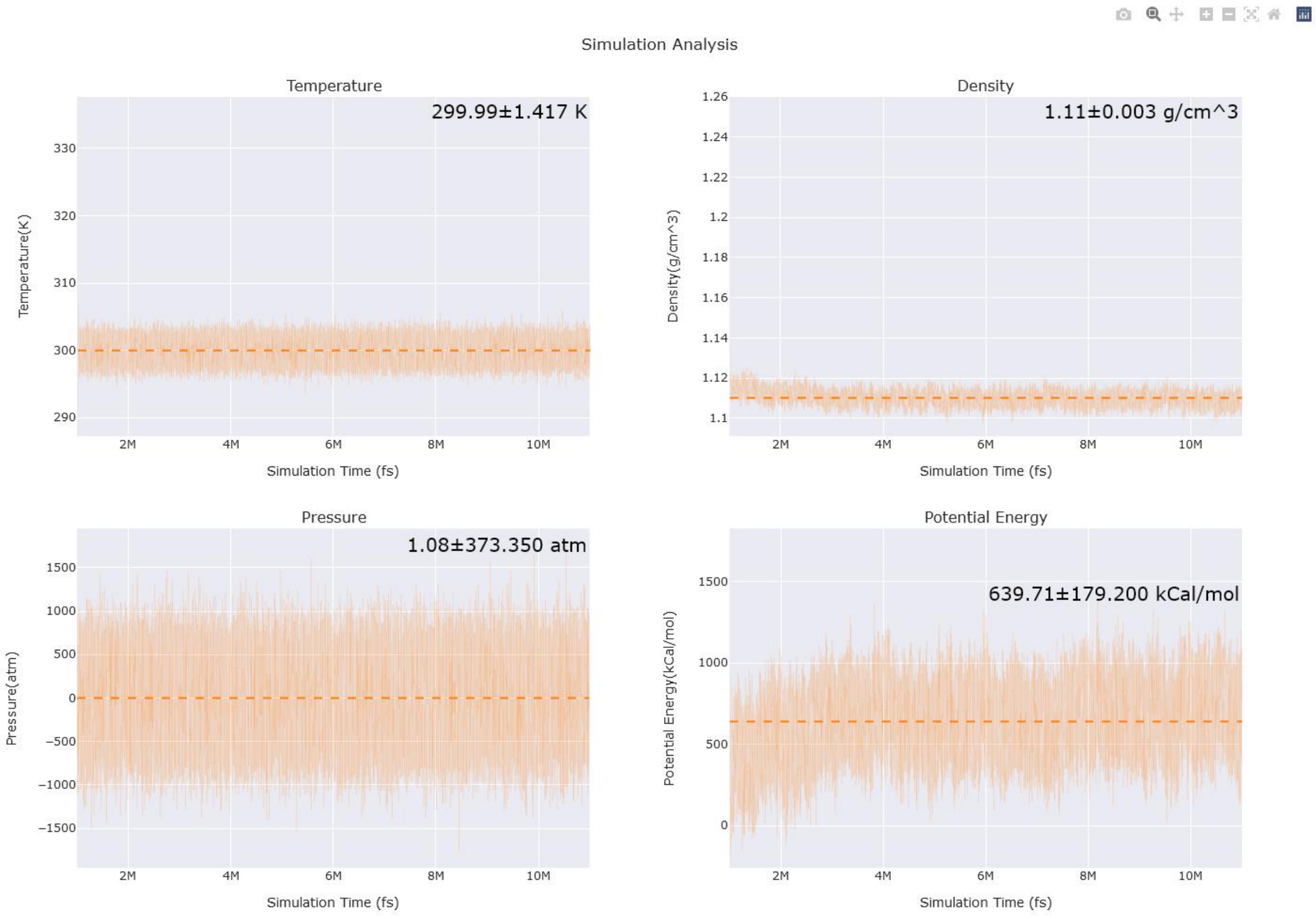

**Figure 3.** Simulation analysis: temperature vs. time (upper left), pressure vs. time (upper right), density vs. time (lower left), potential energy vs. time (lower right). All properties converge in 200 ps.

The measured temperature converges at 300.0 ±1.4 K, density is measured at 1.109±0.035 g/cm$^3$, and total energy is measured at 454±26 kCal/mol. The fluctuations of temperature, energy, and pressure are expected due to the small number of atoms and the duration of the simulation[53]. The calculated density is comparable to an experimental result[54] of 1.124~1.126 g/cm$^3$ with -1.4%~-1.2% error. The simulation is conducted on a workstation with 2 Xeon E5-2690 v4 CPUs. The LAMMPS version is the update 2 of the stable release 23 June 2022, and it is compiled with Intel compilers and Intel MPI. The total simulation CPU time is 66 hours 20 minutes 26 seconds.

The NPT simulation trajectory dumped in both .xyz format and LAMMPS custom format can both be loaded into the Plotly-based visualization tool. The visualization interface is similar to the LAMMPS data file visualizer (Supporting Information 2) but with animation capabilities. The play and pause buttons can control the animation display, and a sliding bar is available for users to wind and rewind the trajectory animation.

The building blocks of the polymer can be edited by accessing the .lmp files. Users can define any building blocks and modify the polymer building loop in order to generate linear polymers. Mixing different monomers can produce copolymer molecules. With one or more three-anchor monomers involved, a nonlinear polymer molecule with branches can also be generated.

*<u>Case-2: preparing a micelle simulation with PEO-PPO copolymers</u>*

Using the building blocks and assembly functionality, a copolymer of polyethylene oxide (PEO) and polypropylene oxide (PPO) can also be built within the PolyMAPS, as shown in Figure 4.

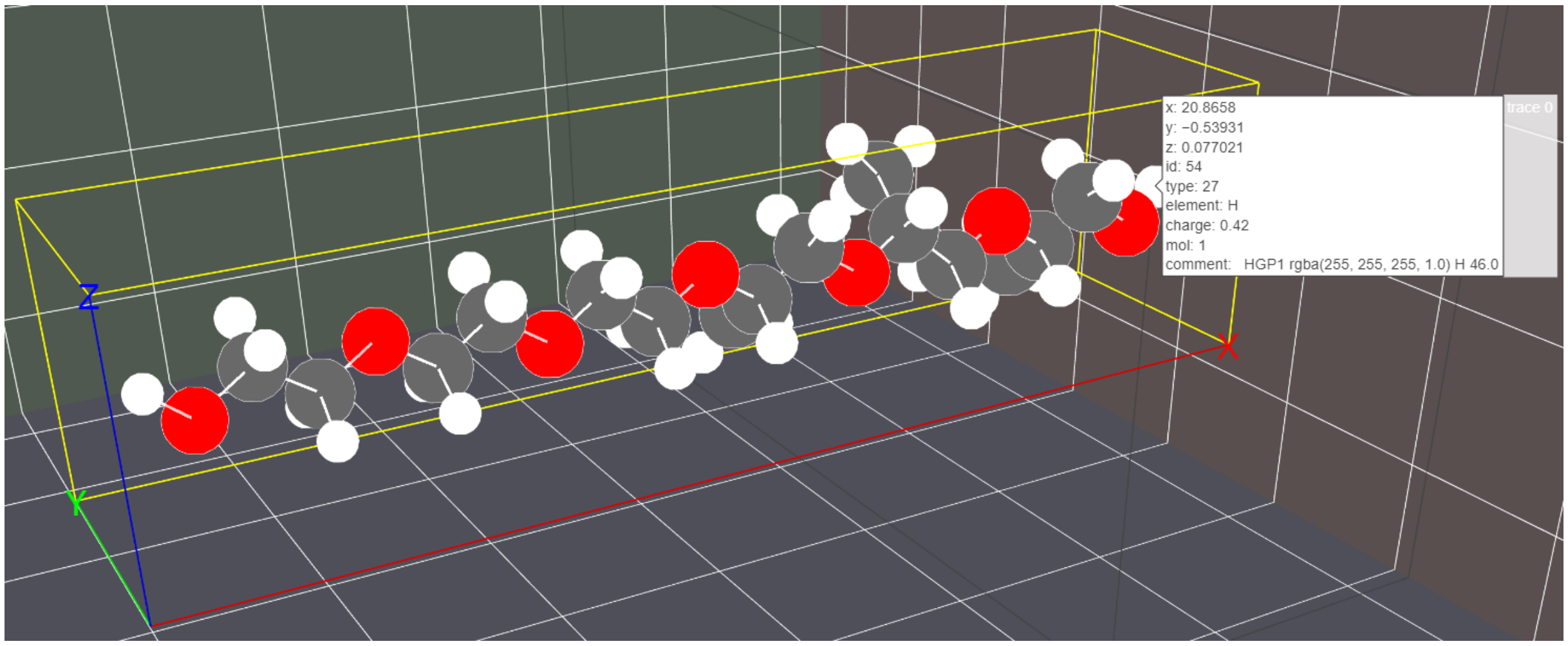

**Figure 4.** A block copolymer chain with 3 ethylene oxide monomers and 3 propylene oxide monomers.

With the help of NumPy, vectorized operations of linear transformations can be applied to the atom coordinates of the polymer chain. By replicating the copolymer, a micelle with copolymer with user-defined orientation can be built, as shown in Figure 5.

The copolymer chain is assigned with CGenFF force field. The interaction types are replicated accordingly during the assembly of the micelle. In order to prepare the simulation, water molecules need to be added to solvate the micelle. The CHARMM flavor of tip3p [55,56] water model is selected. A collision-avoiding program is implemented to prevent the atoms being placed too close to each other. The simulation with both the micelle and the water molecules can be visualized in Figure S2. User has the option to experiment with different block lengths combinations to form a stable micelle in solvent. For example, a stable PEO-PPO copolymer micelle forms around 10~100 nm[57], which is much longer than the example here.

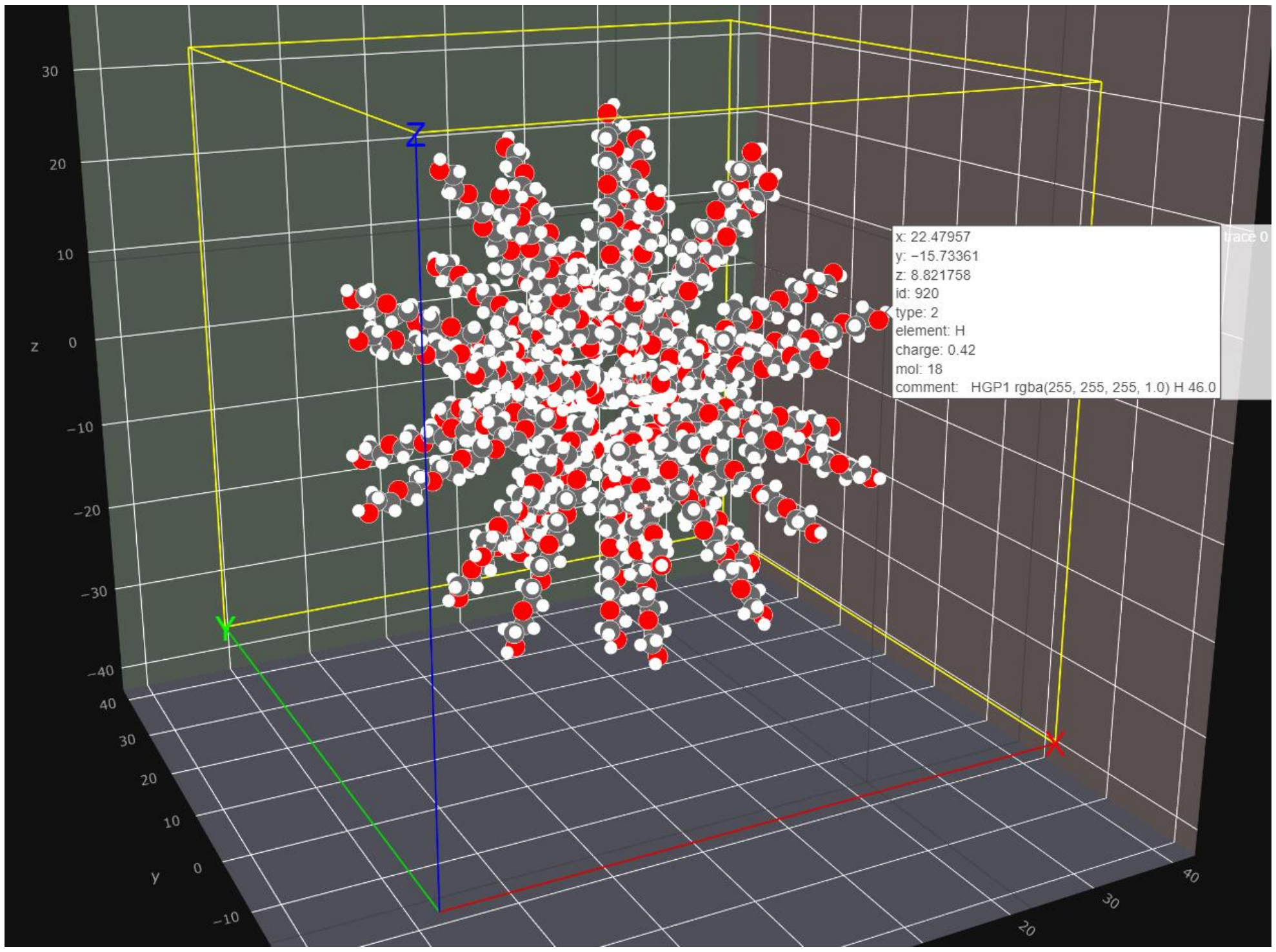

**Figure 5.** A micelle with 36 copolymer chains is built with PolyMAPS.

*Case-3: Viscosity estimation by simulation*

Viscosity calculation by MD simulation is supported by LAMMPS. Here a study of 3 liquid molecules with the Green-Kubo method of viscosity calculation is demonstrated. The Green-Kubo[57,58] method calculates the viscosity by sampling the autocorrelation function of all three off-diagonal stress tensor elements over long time periods. The dynamic viscosity of the fluid can be expressed as:

$$\eta = \frac{V}{k_B T}\int_0^\infty \langle P_{xz}(t_0)P_{xz}(t_0+t)\rangle_{t_0} dt \ ^{59}$$

where $V$ is the volume of the simulation box, $k_B$ is the Boltzmann constant, $T$ is temperature, $P_{xz}$ is the off-diagonal stress tensor element, and $t$ is the simulation time.

The simulation condition is configured as the following: the temperature at 300 K, time step 0.1 fs/step, autocorrelation length 100,000 steps, total simulation time 2,000,000 steps. The molecule is replicated 8,000 times in LAMMPS and equilibrated under an NPT ensemble for more than 10 ns. The production run is conducted under an NVT ensemble. The three liquid molecules under examination are water, cyclohexanone, and 2-(2-Methoxyethoxy)ethanol (DEGME). Each type of liquid molecule is generated by the LigParGen server with the OPLS-AA force field. Water molecule is parameterized as tip4p/2005[60]. The initialized simulation box can be visualized with PolyMAPS as in Figure S1.

The simulation workflow predicts the viscosity values of water, cyclohexanone, and DEGME within a 10% relative error. The results can be summarized in Table 1. Surface tension can also be predicted using the same workflow. The results can be summarized in Table S5.

**Table 1.** Viscosity calculation compared against experimental values; percentage error is included in the parenthesis

| | water (tip4p/2005) | cyclohexanone | DEGME |
|---|---|---|---|
| Experimental Viscosity (mPa•s) at 25°C | 0.854 [61] | 2.2 [62] | 3.48 [63] |
| Green-Kubo Viscosity (mPa•s) at 300 K | 0.79919 (-6.4%) | 2.2461 (+2.1%) | 3.5493 (+2.0%) |

In the table above, column (1) specifies the viscosity value obtained through either experiments or the Green-Kubo simulation method, column(2) is the viscosity value comparison of water, column (3) is the viscosity value comparison of cyclohexanone, and column (4) is the viscosity value comparison of DEGME.

**Conclusion and future work**

PolyMAPS provides researchers with a new approach to initialize simulations in Jupyter notebook with Python language and popular libraries, such as NumPy, SciPy, Pandas, Plotly, and Py3Dmol. The simulation initialization process is separated into 3 parts: atomic coordinates, molecular topology, and force field parameters. Users still reserve the ability to choose whether to use PolyMAPS or other software to process these 3 types of information. PolyMAPS provides an in-place analysis and visualization interface to molecular systems defined in LAMMPS data format. A simple batch job preparation and submission system is provided so that high-throughput simulations are possible in computational clusters. Users can now look up atomic and bonding information directly in the 3D interactive plot. PolyMAPS has significantly reduced the learning curve and the number of tools for LAMMPS users by integrating all simulation processes into a single Jupyter notebook.

PolyMAPS is still in the early stages of development, so features and capabilities are limited to LAMMPS simulation initialization and partial result analysis. The computational job management system can be improved with existing tools, such as FireWorks[64]. The visualization is developed using the Plot.ly python library inside the Jupyter notebook, so the 3D image rendering quality and interaction responsiveness are limited by the WebGL[65] library. Aesthetically, the WebGL-based rendering engine of Plot.ly does support 3D lighting rendering, but this will only increase the already-too-heavy computation load in the rendering process. Most of the existing visualization tools, like VMD and OVITO, support 3D lighting rendering. In the future, the interactive plotting interface can be improved by leveraging more hardware acceleration with more atoms in the system and better 3D lighting effects on the atoms and bonds.

ASSOCIATED CONTENT

**Supporting Information:**

1. Tables S1~S5, Figure S1~S10, .docx file.

2. Polyethylene oxide simulation trajectory visualized by PolyMAPS, .mp4 file.

Data and Software Availability: The PolyMAPS library can be installed from PyPI: https://pypi.org/project/polymaps. A collection of Jupyter notebooks with examples and tutorials can be found at the GitLab repository: https://gitlab.com/doublylinkedlist/polymaps/-/tree/master/notebooks.

AUTHOR INFORMATION

**Corresponding Author**

Santanu Chaudhuri – Department of Civil, Materials, and Environmental Engineering, University of Illinois at Chicago, 815 West Van Buren St, Suite 525, Chicago, IL 60607, United States; also at Applied Materials, and Data Science and Learning Division, Argonne National Laboratory,

Lemont, Illinois 60439, USA; ORCID: https://orcid.org/0000-0002-4328-2947, Email: santc@uic.edu

**Author**

Xiaoli Yan – Department of Civil, Materials, and Environmental Engineering, University of Illinois at Chicago, 815 West Van Buren St, Suite 525, Chicago, IL 60607, United States; ORCID: https://orcid.org/0000-0002-6512-2338, Email: xyan11@uic.edu

**Author Contributions**

The manuscript was written through the contributions of all authors. All authors have given approval to the final version of the manuscript.

**Funding Sources**

NSF funding from CMMI, Award number 2037026, is acknowledged.

**Notes**

The authors declare no competing financial interest.

ACKNOWLEDGMENT

We would like to thank the division of Civil, Mechanical and Manufacturing Innovation (CMMI) of the National Science Foundation (NSF) for its funding support.

# Supporting Information

# An Interactive Polymer Building Toolkit for Molecular Dynamics Simulations: PolyMAPS

*Xiaoli Yan*[*,1], *Santanu Chaudhuri*[1,2]

[1]Department of Civil, Materials, and Environmental Engineering, University of Illinois at Chicago, 815 West Van Buren St, Suite 525, Chicago, IL 60607, United States

[2]Also at: Applied Materials, and Data Science and Learning Division, Argonne National Laboratory, Lemont, Illinois 60439, USA

*Author to whom correspondence should be addressed: xyan11@uic.edu

**Table S1.** LJ interaction parameters

| | atom type | epsilon | Rmin/2 | eps,1-4 | Rmin/2,1-4 | Comment |
|---|---|---|---|---|---|---|
| 1 | HGA2 | -0.035 | 1.34 | 0 | 0 | # alkane, igor, 6/05 |
| 2 | HGP1 | -0.046 | 0.2245 | 0 | 0 | # polar H |
| 3 | CG321 | -0.056 | 2.01 | -0.01 | 1.9 | # alkane (CT2), 4/98, yin, adm jr, also used by viv |
| 4 | OG301 | -0.1 | 1.65 | 0 | 0 | # ether; LJ from THP, sng 1/06 |
| 5 | OG311 | -0.1921 | 1.765 | 0 | 0 | # og MeOH and EtOH 1/06 (was -0.1521 1.7682) |

In the table above, column (1) is the atomic type index, column (2) is the atom type label from CGenFF, column (3)-(6) are the parameters of the LJ interaction, column (7) is the author's comment.

**Table S2.** Bond interaction parameters

| | at1 | at2 | Kb | b0 | Comment |
|---|---|---|---|---|---|
| 1 | CG321 | CG321 | 222.5 | 1.53 | # PROT alkane update, adm jr., 3/2/92 |
| 2 | CG321 | OG301 | 360 | 1.415 | # diethylether, alex |
| 3 | CG321 | OG311 | 428 | 1.42 | # PROT methanol vib fit EMB 11/21/89 |
| 4 | CG321 | HGA2 | 309 | 1.111 | # PROT alkane update, adm jr., 3/2/92 |
| 5 | OG301 | OG311 | 300 | 1.461 | # PBG, yxu, RNA |
| 6 | OG311 | HGP1 | 545 | 0.96 | # PROT EMB 11/21/89 methanol vib fit; og tested on MeOH EtOH,... |

In the table above, column (1) is the bond type index, column (2)-(3) are the atom type labels involved in the bond interactions from CGenFF, column (4)-(5) are the parameters of the bond interaction, column (6) is the author's comment.

**Table S3.** Angle interaction parameters

| | at1 | at2 | at3 | Ktheta | r | kUB | rUB | comment |
|---|---|---|---|---|---|---|---|---|
| 1 | CG321 | CG321 | CG321 | 58.35 | 113.6 | 11.16 | 2.561 | # PROT alkane update, adm jr., 3/2/92 |
| 2 | CG321 | CG321 | OG301 | 45 | 111.5 | 0 | 0 | # diethylether, alex |
| 3 | CG321 | CG321 | OG311 | 75.7 | 110.1 | 0 | 0 | # PROT MeOH, EMB, 10/10/89 |
| 4 | CG321 | CG321 | HGA2 | 26.5 | 110.1 | 22.53 | 2.179 | # PROT alkane update, adm jr., 3/2/92 |
| 5 | OG301 | CG321 | HGA2 | 45.9 | 108.89 | 0 | 0 | # ETOB, Ethoxybenzene, cacha |
| 6 | OG311 | CG321 | HGA2 | 45.9 | 108.89 | 0 | 0 | # PROT MeOH, EMB, 10/10/89 |
| 7 | HGA2 | CG321 | HGA2 | 35.5 | 109 | 5.4 | 1.802 | # PROT alkane update, adm jr., 3/2/92 |
| 8 | CG321 | OG301 | CG321 | 95 | 109.7 | 0 | 0 | # diethylether, alex |
| 9 | CG321 | OG311 | HGP1 | 50 | 106 | 0 | 0 | # sng mod (qm and crystal data); was 57.5 106 |
| 10 | OG301 | OG311 | HGP1 | 61 | 98.3 | 0 | 0 | # PBG, yxu, RNA |

In the table above, column (1) is the angle type index, column (2)-(4) are the atom type labels involved in the angle interactions from CGenFF, column (5)-(6) are the parameters of the angle interaction, column (7) is the author's comment.

**Table S4.** Dihedral interaction parameters

| | at1 | at2 | at3 | at4 | Kchi | n | delta | comment |
|---|---|---|---|---|---|---|---|---|
| 1 | CG321 | CG321 | CG321 | CG321 | 0.0645 | 2 | 0 | # LIPID alkane, 4/04, jbk (Jeff Klauda) |
| 2 | CG321 | CG321 | CG321 | CG321 | 0.14975 | 3 | 180 | # LIPID alkane, 4/04, jbk |
| 3 | CG321 | CG321 | CG321 | CG321 | 0.09458 | 4 | 0 | # LIPID alkane, 4/04, jbk |
| 4 | CG321 | CG321 | CG321 | CG321 | 0.11251 | 5 | 0 | # LIPID alkane, 4/04, jbk |
| 5 | CG321 | CG321 | CG321 | OG301 | 0.16 | 1 | 180 | # methylpropylether, 2/12/05, ATM |
| 6 | CG321 | CG321 | CG321 | OG301 | 0.39 | 2 | 0 | # methylpropylether |
| 7 | CG321 | CG321 | CG321 | OG311 | 0.195 | 3 | 0 | # PROT alkane update, adm jr., 3/2/92 |
| 8 | CG321 | CG321 | CG321 | HGA2 | 0.195 | 3 | 0 | # LIPID alkanes |
| 9 | OG301 | CG321 | CG321 | OG301 | 0.25 | 1 | 180 | # 1,2 dimethoxyethane, 2/12/05, ATM |
| 10 | OG301 | CG321 | CG321 | OG301 | 1.24 | 2 | 0 | # 1,2 dimethoxyethane |
| 11 | OG301 | CG321 | CG321 | HGA2 | 0.19 | 3 | 0 | # alkane, 4/98, yin and mackerell |
| 12 | OG311 | CG321 | CG321 | HGA2 | 0.195 | 3 | 0 | # PROT alkane update, adm jr., 3/2/92 |
| 13 | HGA2 | CG321 | CG321 | HGA2 | 0.22 | 3 | 0 | # LIPID alkanes |
| 14 | CG321 | CG321 | OG301 | CG321 | 0.57 | 1 | 0 | # 1,2 dimethoxyethane, 2/12/05, ATM |
| 15 | CG321 | CG321 | OG301 | CG321 | 0.29 | 2 | 0 | # 1,2 dimethoxyethane |
| 16 | CG321 | CG321 | OG301 | CG321 | 0.43 | 3 | 0 | # 1,2 dimethoxyethane |
| 17 | HGA2 | CG321 | OG301 | CG321 | 0.284 | 3 | 0 | # diethylether, alex |
| 18 | CG321 | CG321 | OG311 | HGP1 | 1.13 | 1 | 0 | # og ethanol |
| 19 | CG321 | CG321 | OG311 | HGP1 | 0.14 | 2 | 0 | # og ethanol |
| 20 | CG321 | CG321 | OG311 | HGP1 | 0.24 | 3 | 0 | # og ethanol |
| 21 | HGA2 | CG321 | OG311 | HGP1 | 0.18 | 3 | 0 | # og methanol |

In the table above, column (1) is the dihedral type index, column (2)-(5) are the atom type labels involved in the dihedral interactions from CGenFF, column (6)-(8) are the parameters of the dihedral interaction, column (9) is the author's comment.

The above information is extracted by filtering the top_all36_cgenff.rtf file with the given keywords related to the polyethylene oxide molecules. from the CHARMM general force field (CGenFF).[1]

**Table S5.** Surface tension calculation compared against experimental values; percentage error is included in the parenthesis.

| | Water (tip4p/2005) | Ethyl Lactate | DEGME |
|---|---|---|---|
| Experimental Surface Tension (mN/m) at 20°C | 72.8[2] | 29.20[3] | 34.8[4] |
| Calculated Surface tension (mN/m) at 293 K | 66.55 (-8.6%) | 28.96 (-0.8%) | 34.93 (+0.4%) |

In the table above, column (1) specifies the surface tension value obtained through either experiments or the MD simulation, column (2) is the surface tension value comparison of water, column (3) is the surface tension value comparison of Ethyl Lactate, and column (4) surface tension value comparison of DEGME.

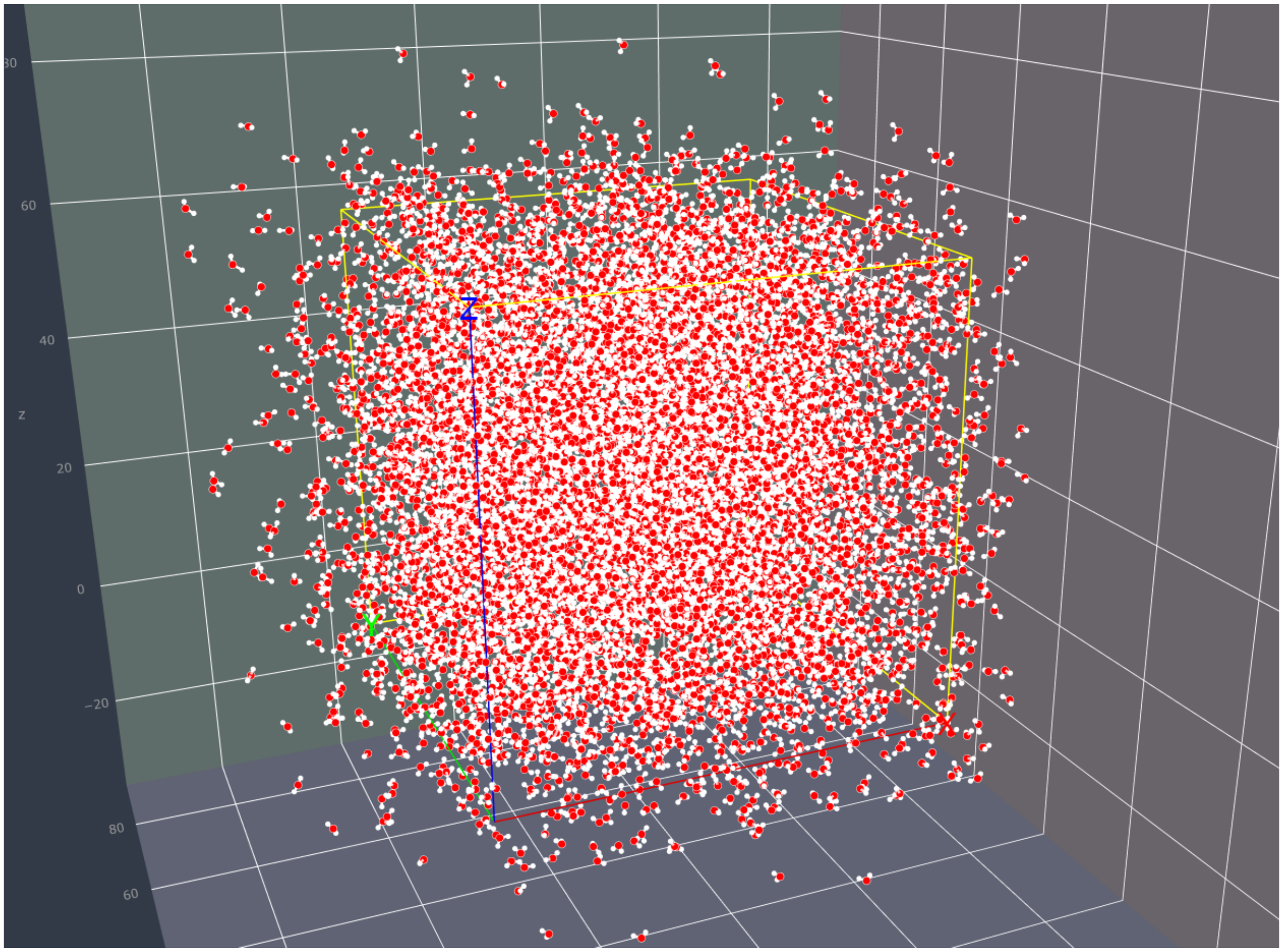

**Figure S1.** Simulation system with 8,000 water molecules in an orthogonal box of size 62 Å×62 Å×62 Å. Given the data point number limitation of Plot.ly, rendering 3D interactive plots with more than 10,000 atoms can be slow depending on the locally available computation resource.

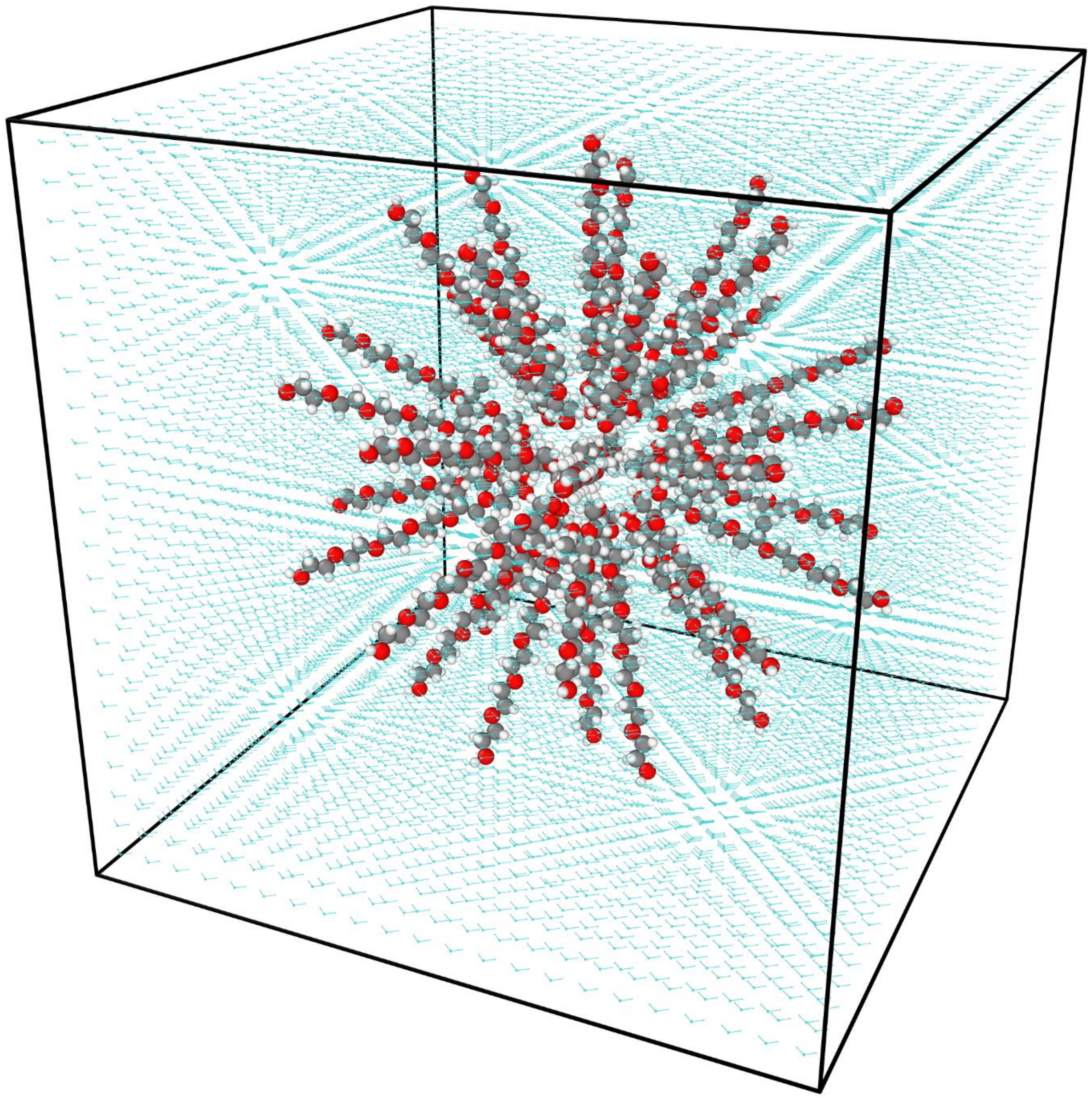

**Figure S2.** Simulation system with a micelle of 36 PEO3-PPO3 copolymer chains and 9,647 water molecules (visualized by OVITO) [5]

*Case-4: Surface tension calculation*

A surface tension calculation analysis tools is included in the PolyMAPS. User can prepare a simulation system with a slab of liquid molecule, as shown in Figure S3. The input script is adapted from the LAMMPS tutorial[6].

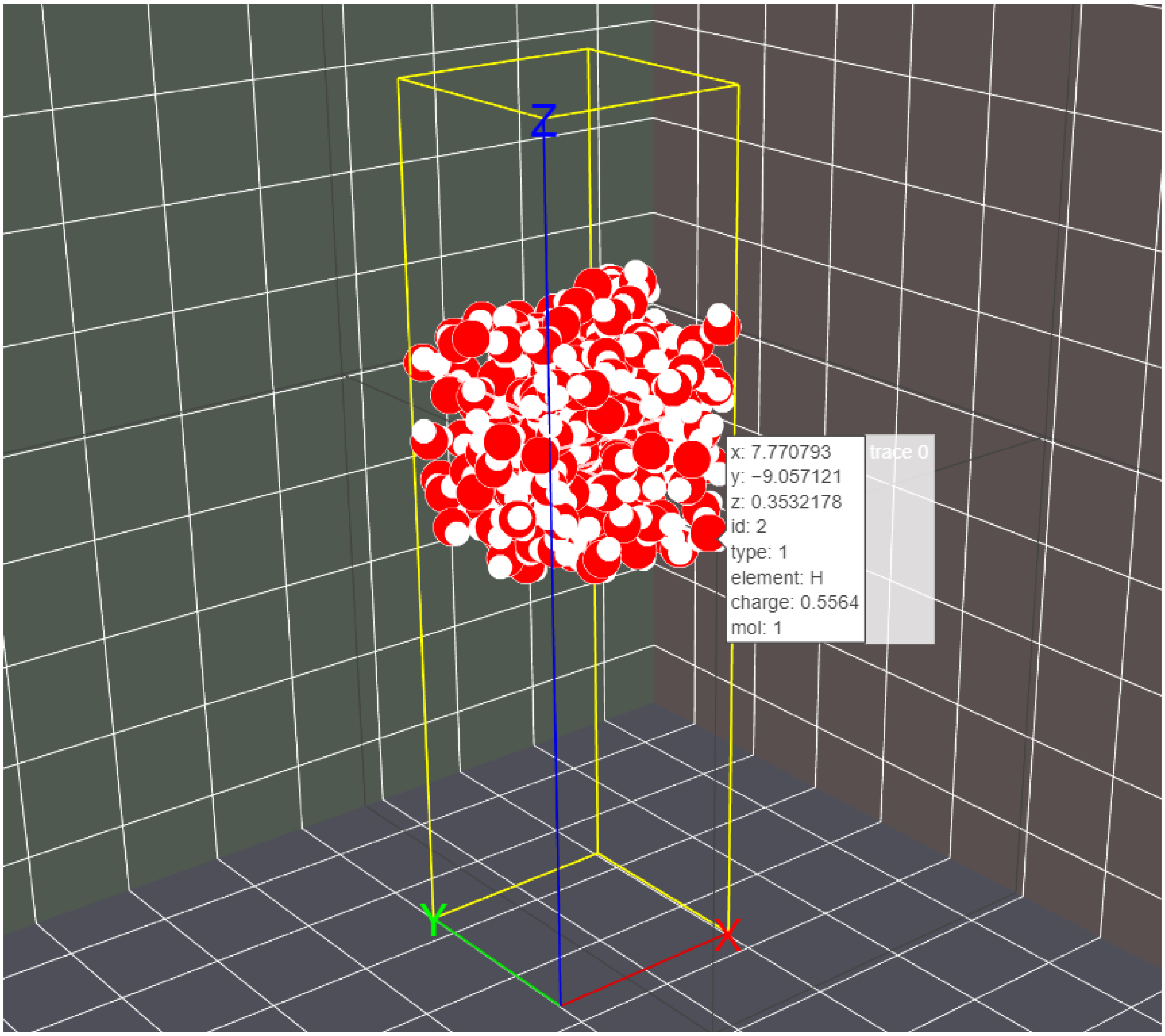

**Figure S3.** Simulation system with a slab of water molecules

The simulation outputs the stress profile in the z-direction vs. time, which can be visualized in Figure S4.

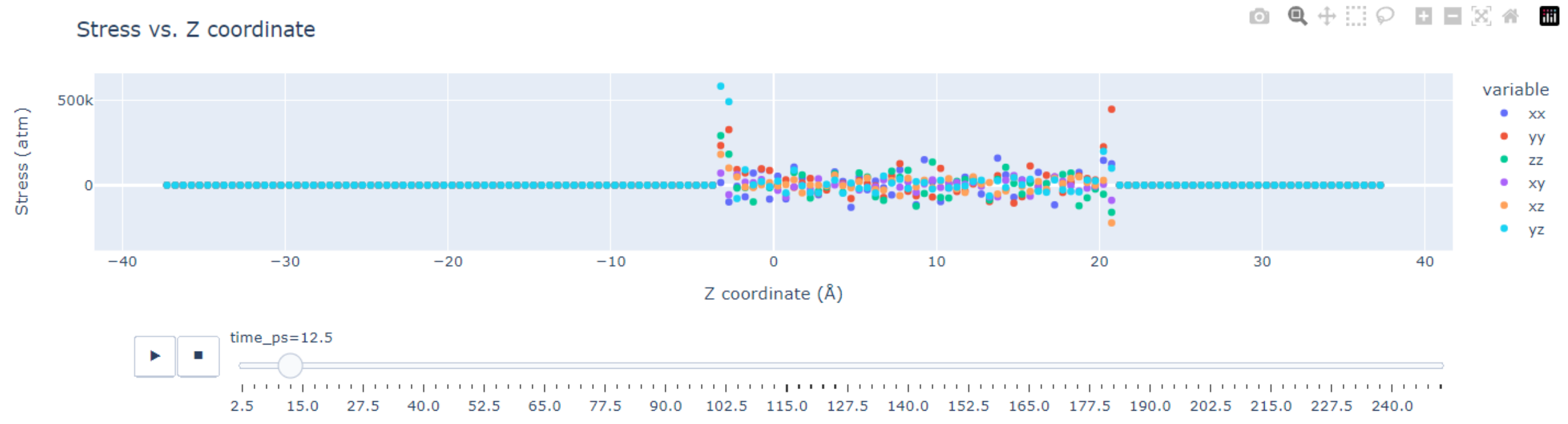

**Figure S4.** Animated visualization of stress profile vs. z coordinate

The calculated surface tension can be visualized as Figure S5.

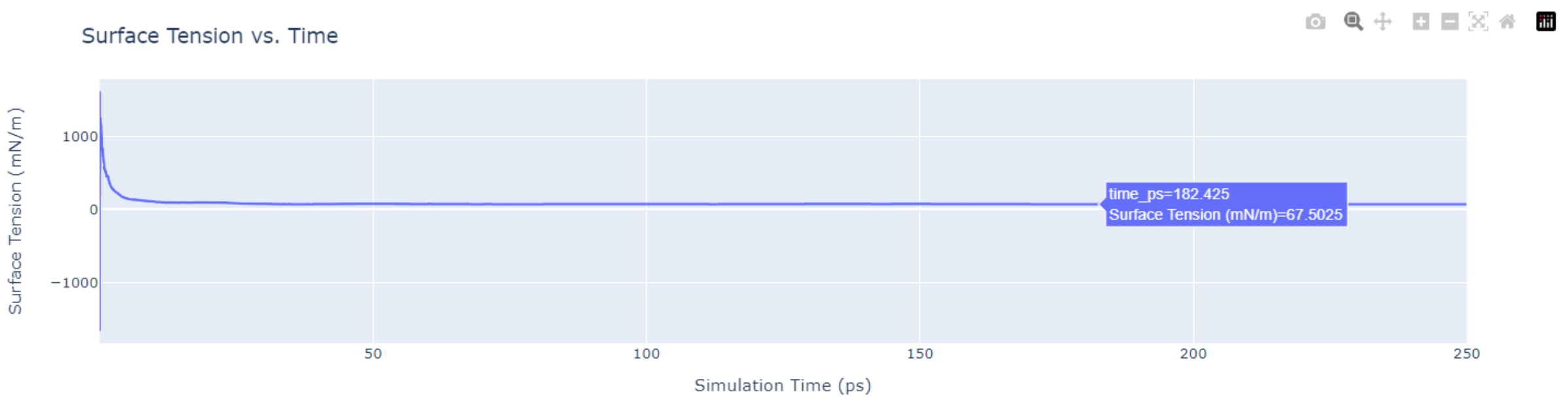

**Figure S5.** Visualization of surface tension vs. simulation time

*Case-5: Reactive polymer building with non-reactive force field*

There is an existing tool called AutoMapper[7], and it provides a more systematic approach to the REACTER[8,9] package in LAMMPS. This part of the PolyMAPS provides a user-friendly and visualization-based approach. Data files and reaction template files before and after the reaction can all be visualized as shown in Figure S6 and S7.

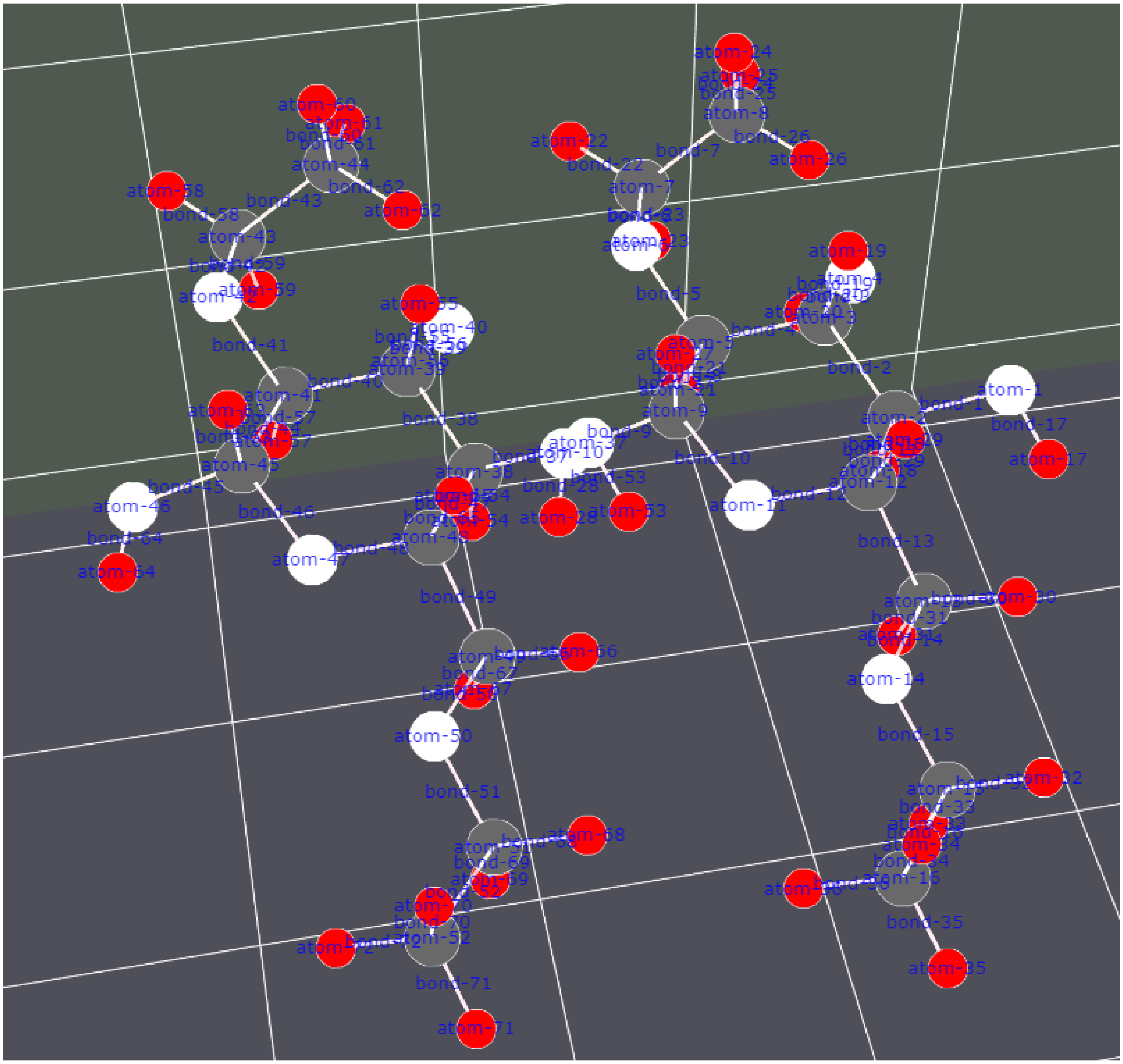

**Figure S6.** Visualization of two ethyl cellulose (EC) monomer units before reaction

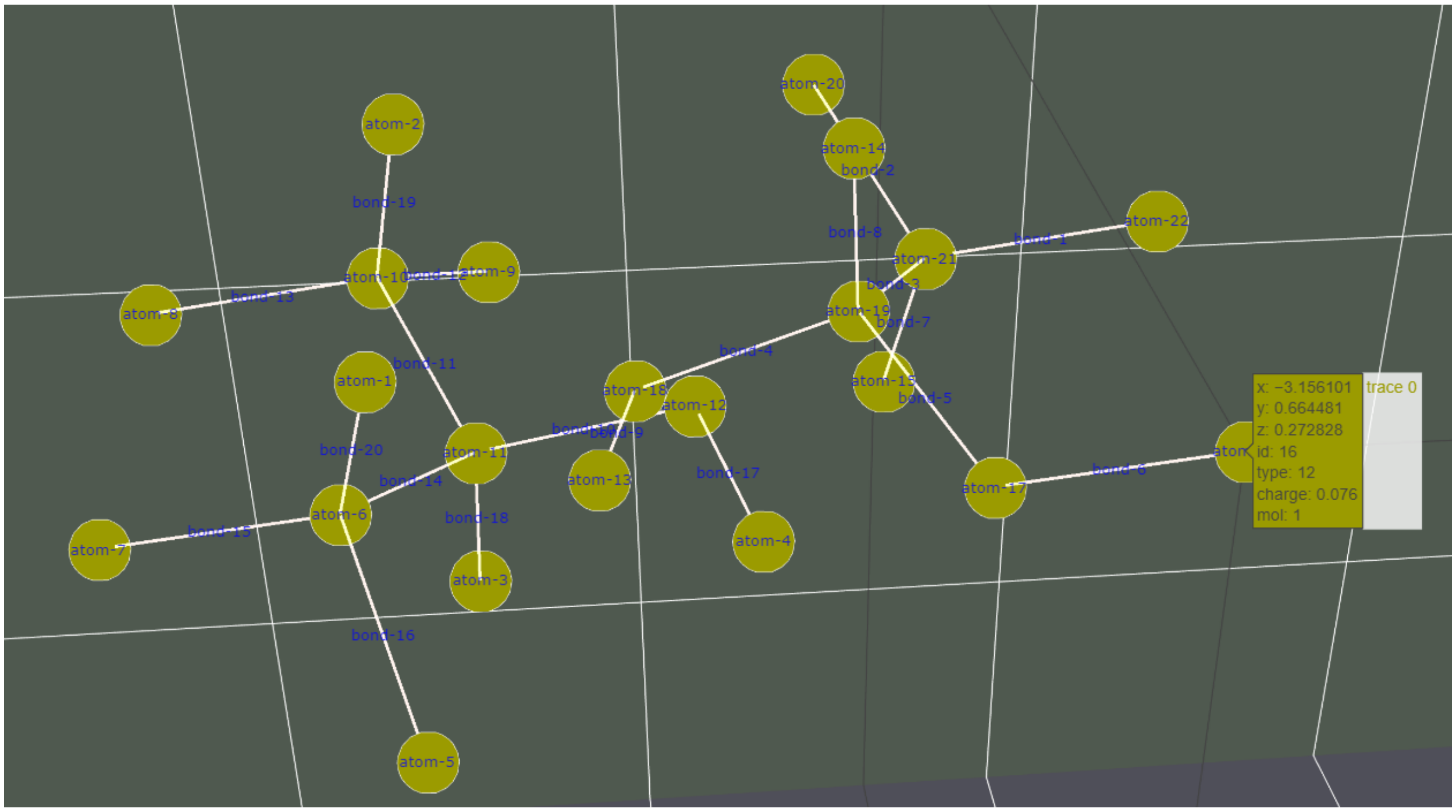

**Figure S7.** Visualization of the reaction template file of the two ethyl cellulose (EC) monomer units before reaction

The reaction template file is automatically generated once the initiator atoms are selected by the user. The reaction counter can be visualized in Figure S8.

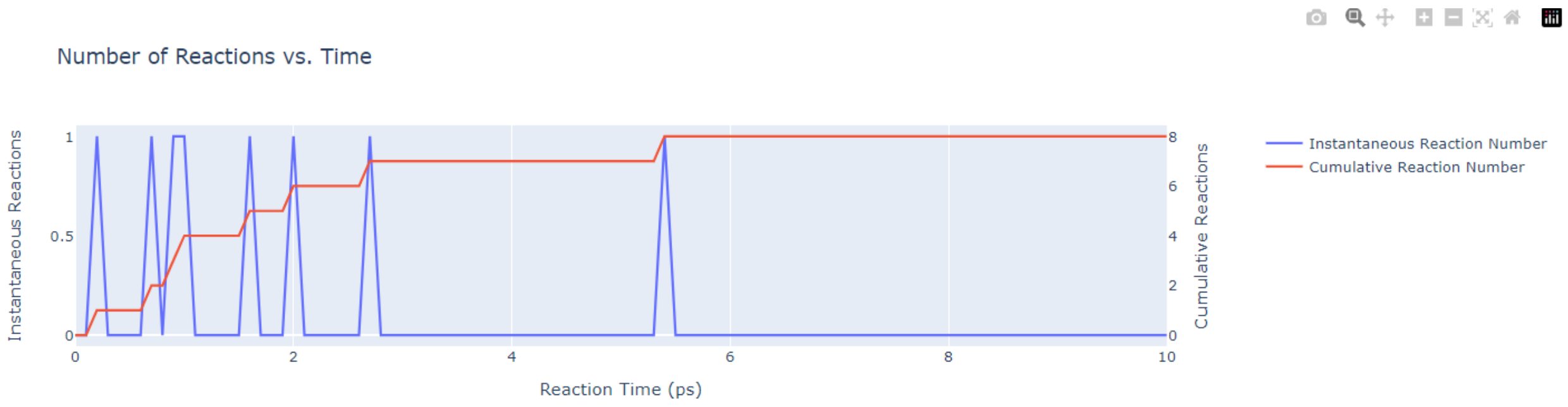

**Figure S8.** Visualization of the reaction counter vs. simulation time

The atoms in the reacted system can be visualized with color-coded based on the molecule id, as shown in Figure S9.

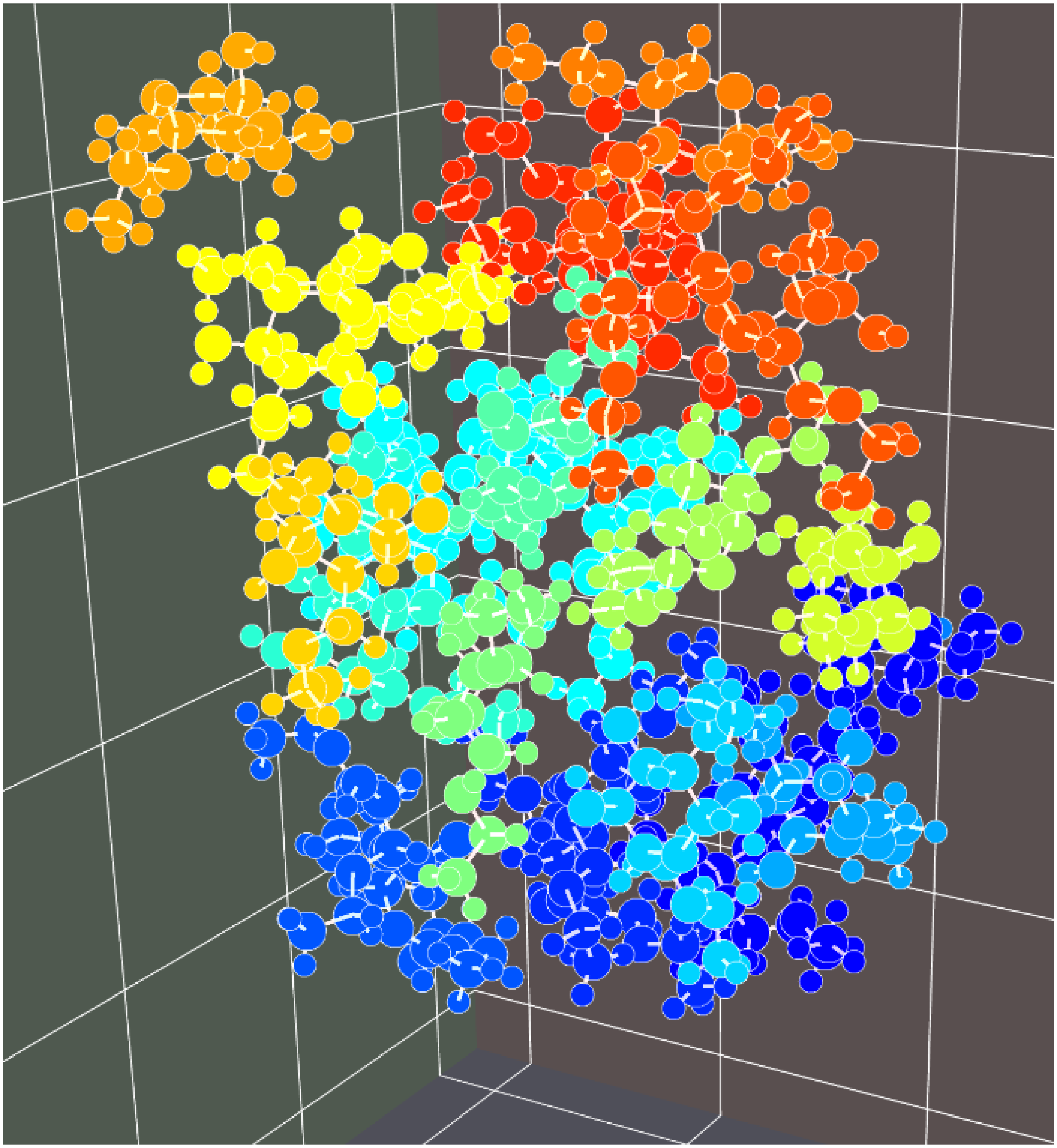

**Figure S9.** Visualization of final state of the reactive simulation on 27 EC monomer units, different color denotes different molecule ids.

<u>*Case-6: Concentration-density profile of Isopropanol and water mixture*</u>

A series of NPT simulations of isopropanol-water mixture at different concentrations are conducted with OPLS-AA[10,11] force field and TIP4P/2005[12] water model. Using PolyMAPS's built-in batch job scripting function, all simulation files can be prepared with both Torque and Slurm job scripts. The simulation results are compared against experimental results[13]. The code is available as an example notebook in repository. The density vs. concentration profile is demonstrated in Figure S10.

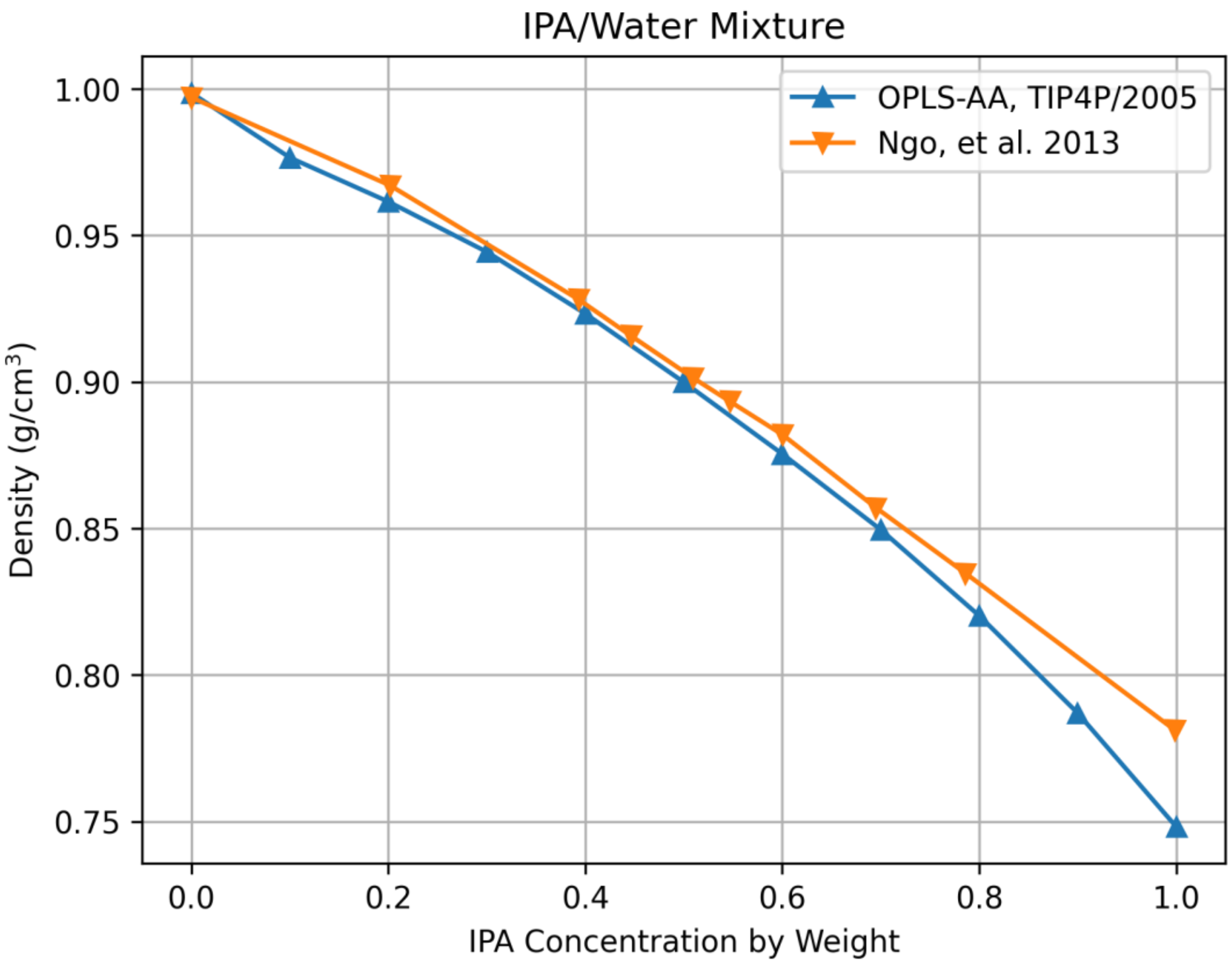

**Figure S10.** Visualization of final state of the reactive simulation on 27 EC monomer units, different color denotes different molecule ids.

**Table of Content graphic**

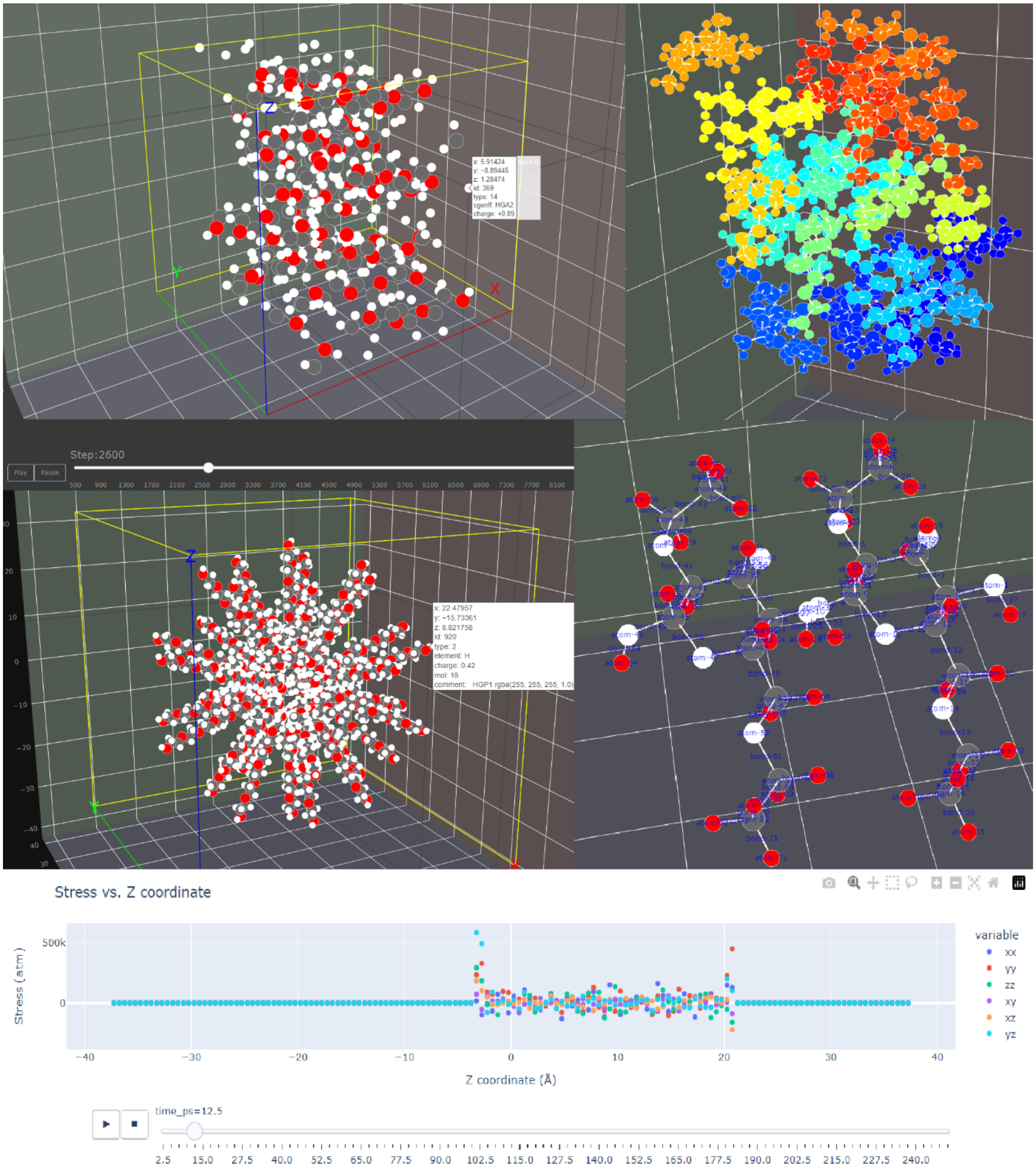